\documentclass[aps,twocolumn,reprint,english, superscriptaddress,showpacs,longbibliography]{revtex4-1}

\usepackage{graphicx}
\usepackage{subfigure}
\usepackage{dcolumn}
\usepackage{bm}
\usepackage{amsmath}
\usepackage{amssymb}
\usepackage{latexsym}
\usepackage{epsfig}
\usepackage{amsbsy}
\usepackage{array}
\usepackage{amssymb}
\usepackage{subeqnarray}
\usepackage{cases}

\usepackage{setspace}
\usepackage{color}
\usepackage{ulem}

\definecolor{darkblue}{rgb}{0.1,0.2,0.6}
\definecolor{darkred}{rgb}{0.8,0.1,0.2}
\usepackage[colorlinks,
            citecolor=darkblue,
            linkcolor=darkred,
            urlcolor=darkblue]{hyperref}
\usepackage[all]{hypcap}
\usepackage{braket}

\begin{document}

\makeatletter
\adddialect\l@English\l@english
\makeatother

\newcommand{\br}{\mathbf{r}}
\newcommand{\brprime}{{\mathbf{r}^\prime}}
\newcommand{\bk}{\mathbf{k}}
\newcommand{\bkprime}{{\mathbf{k}^\prime}}
\newcommand{\bq}{\mathbf{q}}
\newcommand{\hx}{\hat{x}}
\newcommand{\hy}{\hat{y}}

\newcommand{\up}{\uparrow}
\newcommand{\dn}{\downarrow}
\newcommand{\la}{\langle}
\newcommand{\ra}{\rangle}


\title{Single-orbital realization of high-temperature $s^\pm$ superconductivity in the square-octagon lattice }
\author{Yao-Tai Kang}
\thanks{These two authors contributed equally to this work.}
\affiliation{State Key Laboratory of Optoelectronic Materials and Technologies, School of Physics, Sun Yat-sen University, Guangzhou 510275, China}
\affiliation{Department of Physics, National Tsing Hua University, Hsinchu 30013, Taiwan}
\author{Chen Lu}
\thanks{These two authors contributed equally to this work.}
\affiliation{School of Physics, Beijing Institute of Technology, Beijing 100081, China}
\author{Fan Yang}
\email{yangfan\_blg@bit.edu.cn}
\affiliation{School of Physics, Beijing Institute of Technology, Beijing 100081, China}
\author{Dao-Xin Yao}
\email{yaodaox@mail.sysu.edu.cn}
\affiliation{State Key Laboratory of Optoelectronic Materials and Technologies, School of Physics, Sun Yat-sen University, Guangzhou 510275, China}

\begin{abstract}
  We propose possible high-temperature superconductivity (SC) with singlet $s^\pm$-wave pairing symmetry in the single-orbital Hubbard model on the square-octagon lattice with only nearest-neighbor hopping terms. Three different approaches are engaged to treat with the interacting model for different coupling strengths, which yield consistent result for the $s^\pm$ pairing symmetry. We propose octagraphene, i.e., a monolayer of carbon atoms arranged into this lattice, as a possible material realization of this model. Our variational Monte Carlo study for the material with realistic coupling strength yields a pairing strength comparable with the cuprates, implying a similar superconducting critical temperature between the two families. This study also applies to other materials with similar lattice structure.
\end{abstract}
\maketitle

\section{Introduction}
The search for superconductivity (SC) with high critical temperature $T_c$ has been the dream of the condensed-matter community for decades. It is generally believed that the right route to seek for high-$T_c$ SC (HTCS) is to acquire strong spin fluctuations via proximity to antiferromagnetic-ordered phases, with the cuprates and the iron-based superconductors as two well-known examples \cite{DJScalapino12}. Along this route, a new research area was generated recently:  graphene-based SC. Among the early attempts in this area, the most famous idea might be to generate d+id HTCS \cite{Chubukov,Qianghua,Thomale} in the monolayer graphene in proximity to the spin-density-wave (SDW) ordered state \cite{TaoLi,Qianghua} at the quarter-doping. However, such high doping concentration is hardly accessible by experiment. The newly discovered SC in the magic-angle-twisted bilayer graphene \cite{Cao1} in close proximity to the ``correlated insulator" phase \cite{Cao2} opened a new era in this area. It is proposed that the ``correlated insulator" in this material is a SDW insulator \cite{Yang, Xu}, and the SC is driven by SDW spin fluctuations \cite{Yang,Xu,Ashvin,Fu}. However, due to the greatly reduced Fermi energy ($\approx10$ meV) in this material, the $T_c\approx 1.7$ K might be not far from its upper limit. Here we propose another graphene-based material, i.e., octagraphene \cite{Sugang}, which has a square-octagon lattice structure with each site accommodating one single $2p_z$ orbital. This system has large Fermi energy and we predict that slightly doping this material will induce HTCS, driven by SDW spin-fluctuations.

The octagraphene is a two-dimensional (2D) material formed by a monolayer of carbon atoms arranged into a square-octagon lattice as shown in Fig.~\ref{fig:lattice}. This lattice is $C_{4v}$-symmetric and each unit cell contains four sites forming a square enclosed by the dotted lines shown in Fig.~\ref{fig:lattice}. First-principles calculations indicate that such a planar structure is kinetically stable at low temperature \cite{Sugang,Pod} and that its energy is a local minimum \cite{Sugang}, which suggests that the material can potentially be synthesized in laboratories. Actually, this lattice structure has attracted a lot of research interest recently because it not only is hosted by quite a few real materials \cite{CaVO,KFeSe1,KFeSe2,YZhang} but also has various intriguing phases on this lattice that have been revealed by theoretical calculations  \cite{Scalettar,Troyer,White,Sachdev,Zheng_Weihong,Bose,Manuel,Farnell,Kwai,Fiete,Yamashita,Yanagi,Yamada14,Wu,Iglovikov,Long_Zhang,Gong,Bao}. Here we notice another remarkable property of this 2D lattice: its band structure can have perfect Fermi-surface (FS) nesting in a wide parameter regime at half filling, which easily leads to antiferromagnetic SDW order. When the system is slightly doped, the SDW order will be suppressed and the remnant SDW fluctuation will mediate HTCS.

In this paper, we study a possible pairing state in the single-orbital Hubbard-model on the square-octagon lattice with only nearest-neighbor hopping terms.  To treat this Hubbard-model with different limits of the coupling strength, we adopt three distinct approaches, i.e., the random-phase approximation (RPA), the slave-boson mean field (SBMF), and the variational Monte Carlo (VMC), which are suitable for the weak, the strong, and the intermediate coupling strengths, respectively. All the three approaches consistently identify the single $s^\pm$-wave pairing as the leading pairing symmetry. We propose octagraphene as a possible material realization of the model. Our VMC calculation adopting realistic interaction strength yields a pairing gap amplitude of about 50 meV, which is comparable with the cuprates, implying a comparable $T_c$ between the two families. Our study also applies to other materials with similar lattice structure.

\section{Material, Model, and Approaches}
\begin{figure}
    \begin{center}
    	\subfigure[]{\includegraphics[width=1.6in]{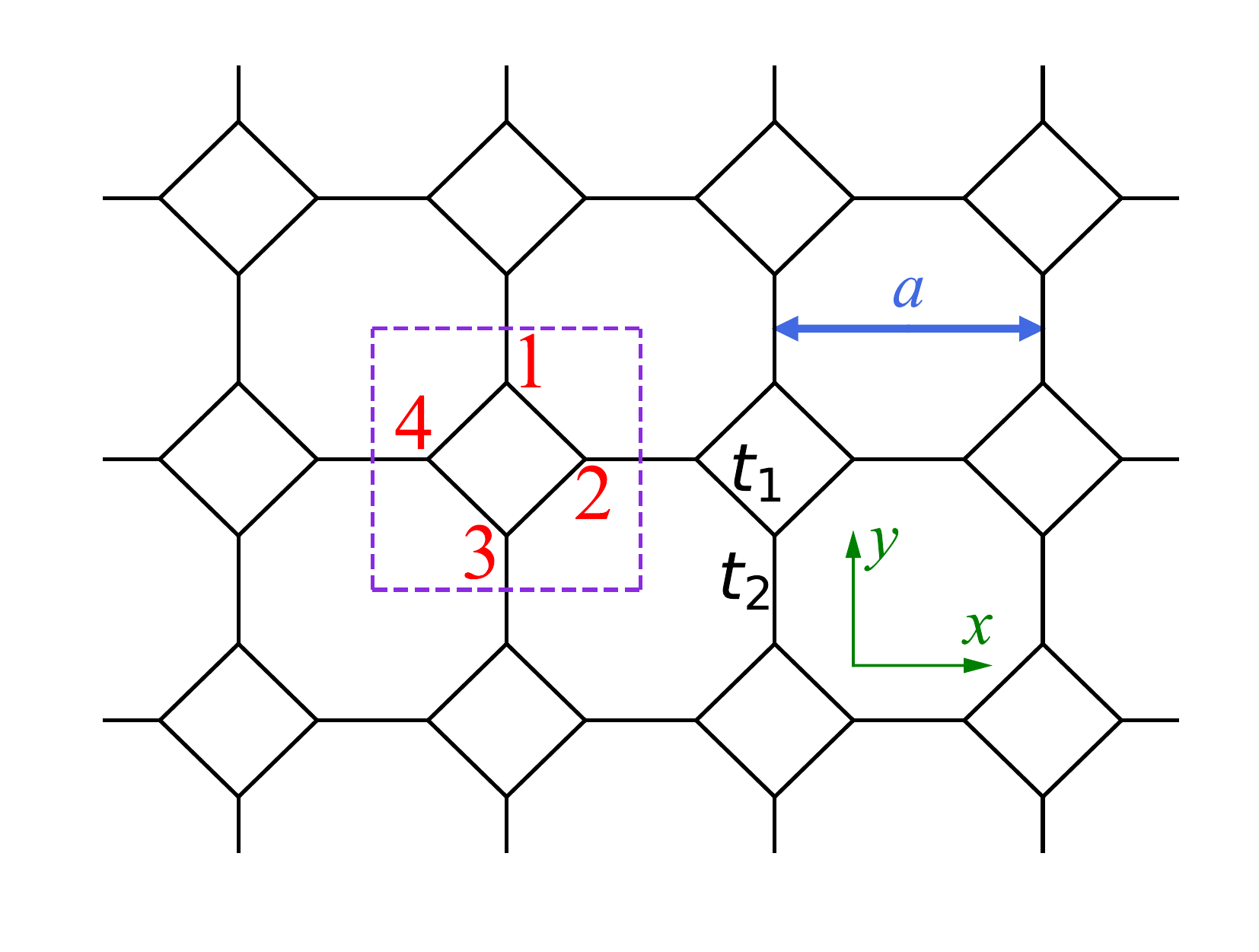}\label{fig:lattice}}
    	\subfigure[]{\includegraphics[width=1.5in]{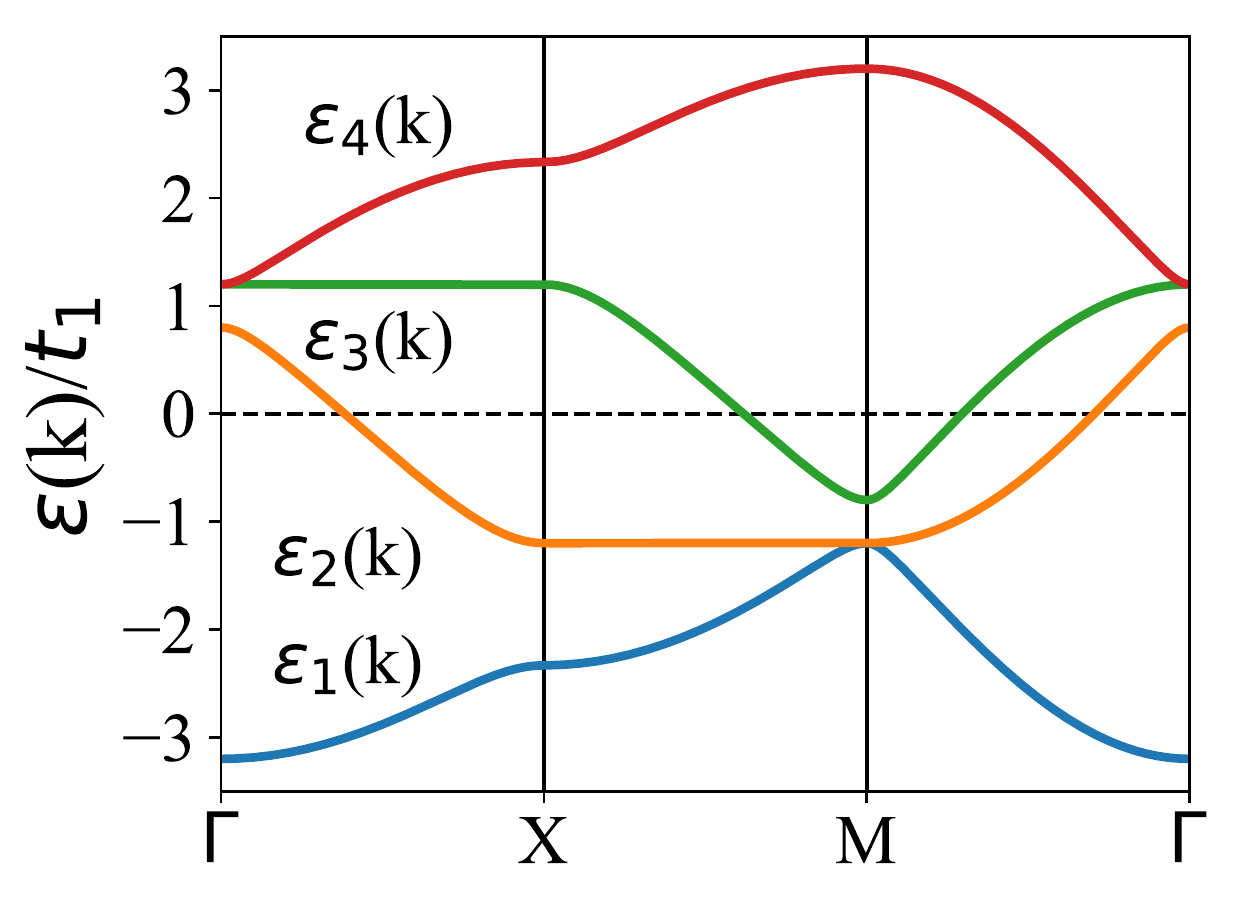}\label{fig:band}}
    	\\
    	\subfigure[]{\includegraphics[width=1.55in]{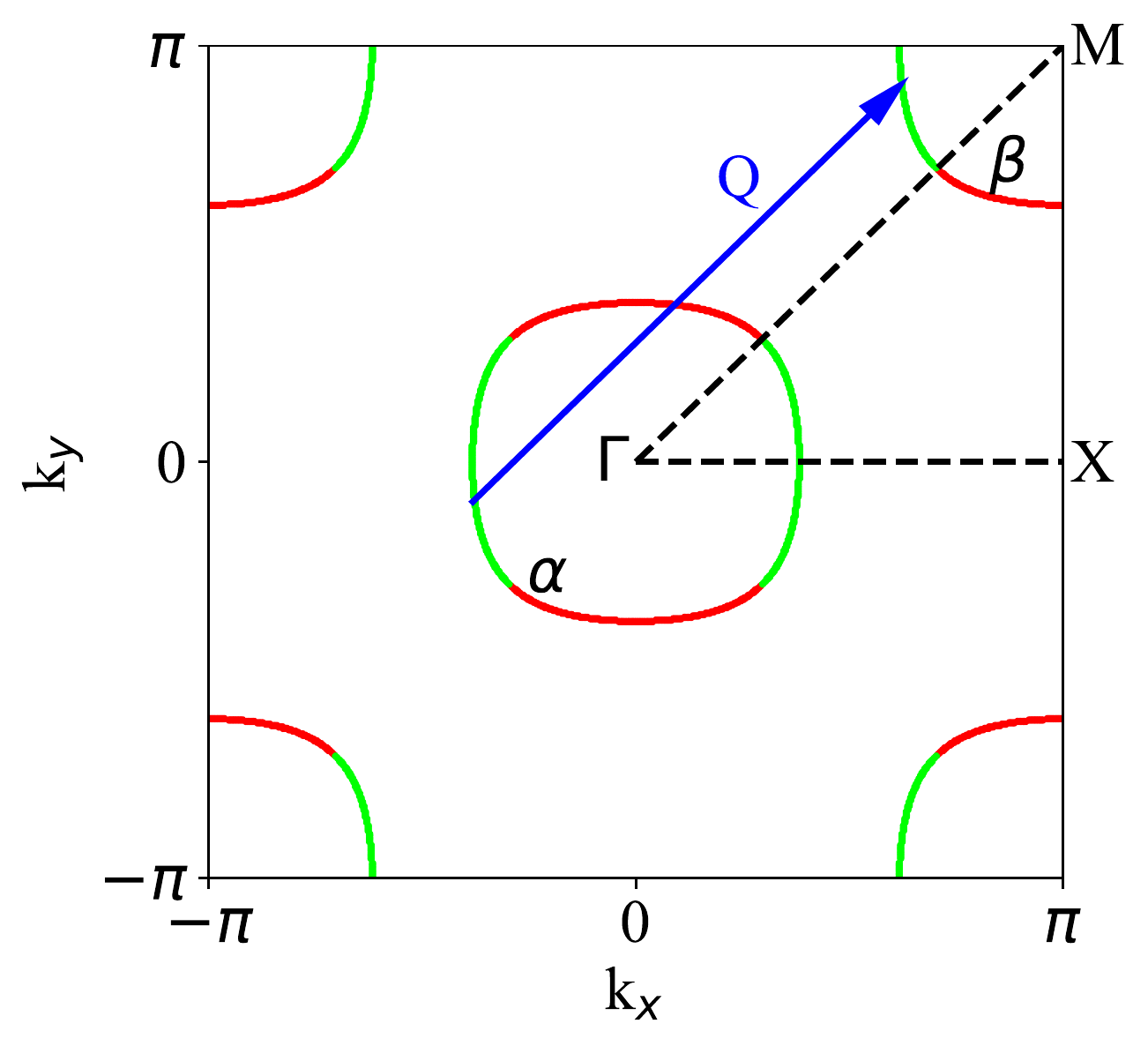}\label{fig:FSn100}}
    	\subfigure[]{\includegraphics[width=1.5in]{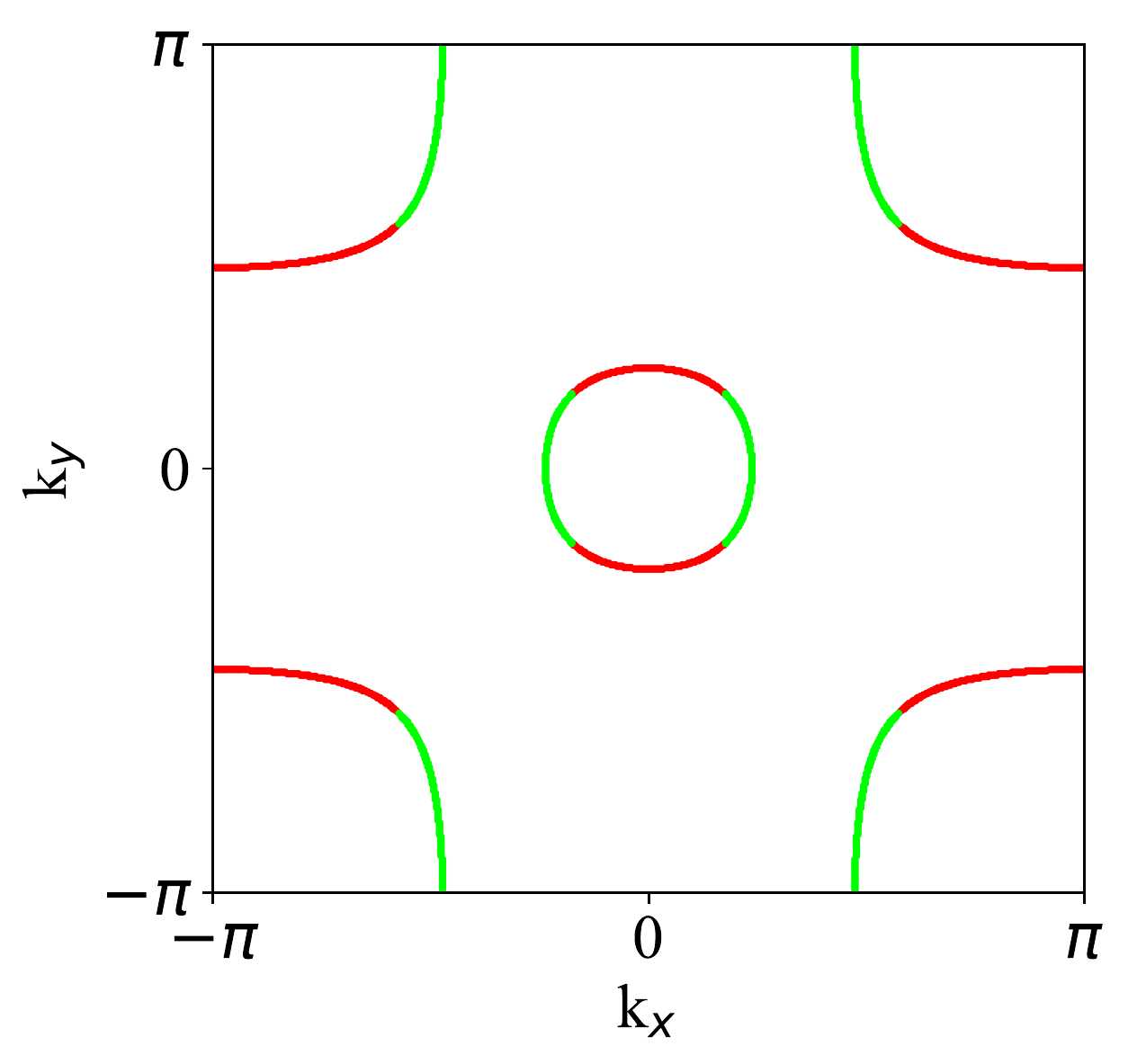}\label{fig:FSn110}}
    	\caption{(a) Sketch of the square-octagon lattice and illustration of the intrasquare nearest-neighbor hopping $t_1$ and the intersquare nearest-neighbor hopping $t_2$. The dotted square denotes the unit cell. (b) Band structure of the TB model (\ref{TB}) along the high symmetric lines in the first Brillouin zone. Panels (c) and (d) show the FSs of the undoped and $10\%$ electron-doped cases, respectively. The site contributions on the FS sheets are shown by color: the red (green) represents that the weights contributed by the sublattices 1 and 3 (2 and 4) are dominant. The TB parameters are $t_1=1,t_2=1.2$ through out the work.}
    \end{center}
\end{figure}
From density-functional theory (DFT) calculations \cite{Sugang}, each carbon atom in the octagraphene is $\sigma$ bonded with its three surrounding atoms via $sp^2$ hybridization. The low-energy degree of freedom near the Fermi level is dominantly contributed by the $2p_z$ orbitals, which form $\pi$ bonds similar to the graphene. With each carbon atom contributing one electron in one $2p_z$ orbital, the resulting band structure can be well captured by the following single-orbital TB model:
\begin{eqnarray}
H_{\text{TB}}=-t_1 \sum_{\langle i,j \rangle,\sigma} \left( c^{\dagger}_{i\sigma} c_{j\sigma} + H.c. \right)
       - t_2 \sum_{[i,j],\sigma} \left( c^{\dagger}_{i\sigma} c_{j\sigma} + H.c. \right).\nonumber\\\label{TB}
\end{eqnarray}
Here $c^{\dagger}_{i\sigma} \left(c_{i\sigma}\right)$ creates (annihilates) an electron with spin $\sigma$ at site $i$.
The terms with coefficients $t_1$ ($\approx$2.5eV) and $t_2$ ($\approx$2.9eV)  describe the intrasquare nearest-neighbor ($NN$) and intersquare $NN$ hoppings respectively, as shown in Fig.~\ref{fig:lattice}. In the following, we set $t_1$ as the energy unit and $t_2/t_1=1.2$.

The band structure of this TB model along the high symmetric lines in the first Brillouin zone is presented in Fig.~\ref{fig:band}.
For the half-filling case, the band $\varepsilon_2(\bk)$ and $\varepsilon_3(\bk)$ cross the Fermi level to form a hole pocket ($\alpha$) centering around the $\Gamma$ point, and an electron pocket ($\beta$) centering around the $M$ point, as shown in Fig.~\ref{fig:FSn100}.
The red (green) color indicates that site 1 and 3 (2 and 4) dominate the weights of bands.
Remarkably, the two pockets are identical, connected by the perfect nesting vector $\mathbf{Q}=(\pi,\pi)$.
Such perfect FS-nesting is robust at half filling in the parameter regime $0 < \left|\frac{t_2}{t_1}\right| \le 2$, where the FS exists. However, upon doping, the perfect FS nesting is broken, leaving a remnant nesting at a nesting vector shifted from $\mathbf{Q}$, as shown in Fig.~\ref{fig:FSn110}.

Due to the screening effect in the doped compound, the strong Coulomb repulsions between the $2p_z$ electrons in the graphene-based material can be approximated as the Hubbard interaction \cite{Neto}. Therefore, we obtain the following well-known (repulsive) Hubbard-model:
\begin{equation}
H =H_{\text{TB}}+H_{\text{int}}=H_{\text{TB}}+U \sum_i \hat{n}_{i\uparrow} \hat{n}_{i\downarrow},\label{model}
\end{equation}
Although there is a rough estimate of $U\approx10$ eV for the graphene-based material, an accurate value of $U$ is hard to obtain \cite{Neto}. Therefore, in the following, we first engage three different approaches, i.e., the RPA, the SBMF, and the VMC, to treat with the model with different limits of $U$ and check the $U$ dependence of the pairing symmetry. As we shall see, they yield consistent results. Then, we fix $U=10$ eV, and adopt the VMC approach suitable for this $U$ to estimate the $T_c$.

\section{Theoretical solutions and numerical results}

\subsection{Results for the random-phase approximation}
We adopt the standard multi-orbital RPA approach \cite{KKubo07,SGraser09,QLLuo10,TAMaier11,FengLiu13,TXMa14,XXWu15,LDZhang15,HKontani98,HKondo01,KKuroki02}
to treat the weak-coupling limit of the model (\ref{model}).
Strictly speaking, this is an ``intra-unit-cell multisite model'' without orbital degrees of freedom, which is easier because of the absence of an inter-orbital Coulomb interaction and Hund's coupling.
This approach handles the interactions at the RPA level, from which we determine the properties of the magnetism and SC for interactions above or below the critical interaction strength $U_c$, respectively. Generally, the RPA approach only works well for weak-coupling systems.

\begin{figure}
	\flushleft
	\subfigure[]{\includegraphics[width=1.63in]{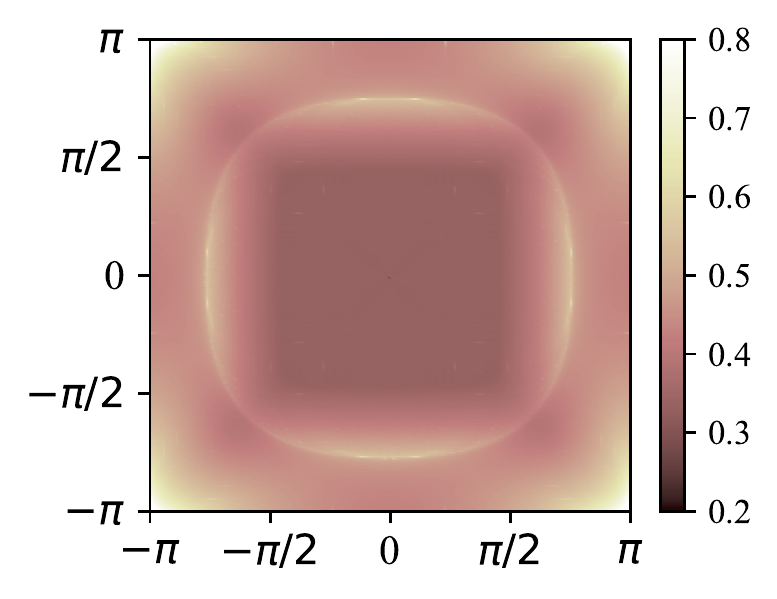}\label{fig:chi0x00}}
	\subfigure[]{\includegraphics[width=1.63in]{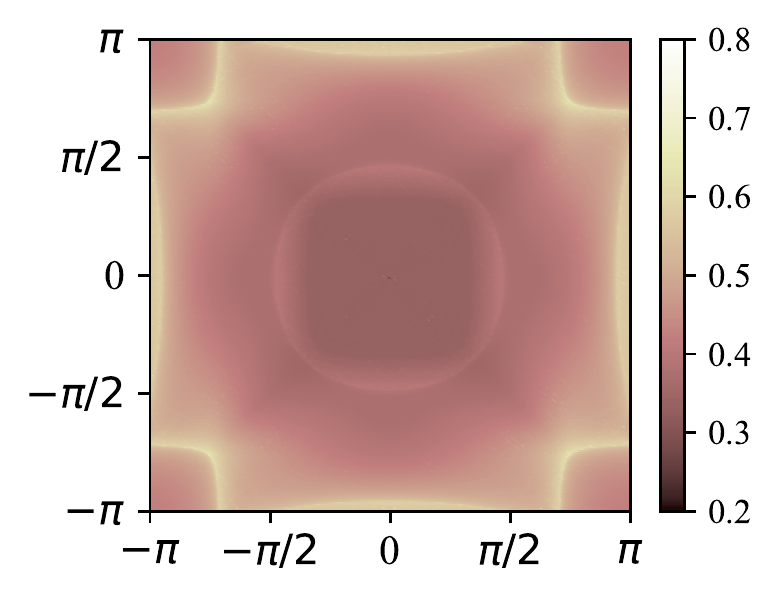}\label{fig:chi0x10}}
	\\
	\hspace{0.02in}
	\subfigure[]{\includegraphics[width=1.35in]{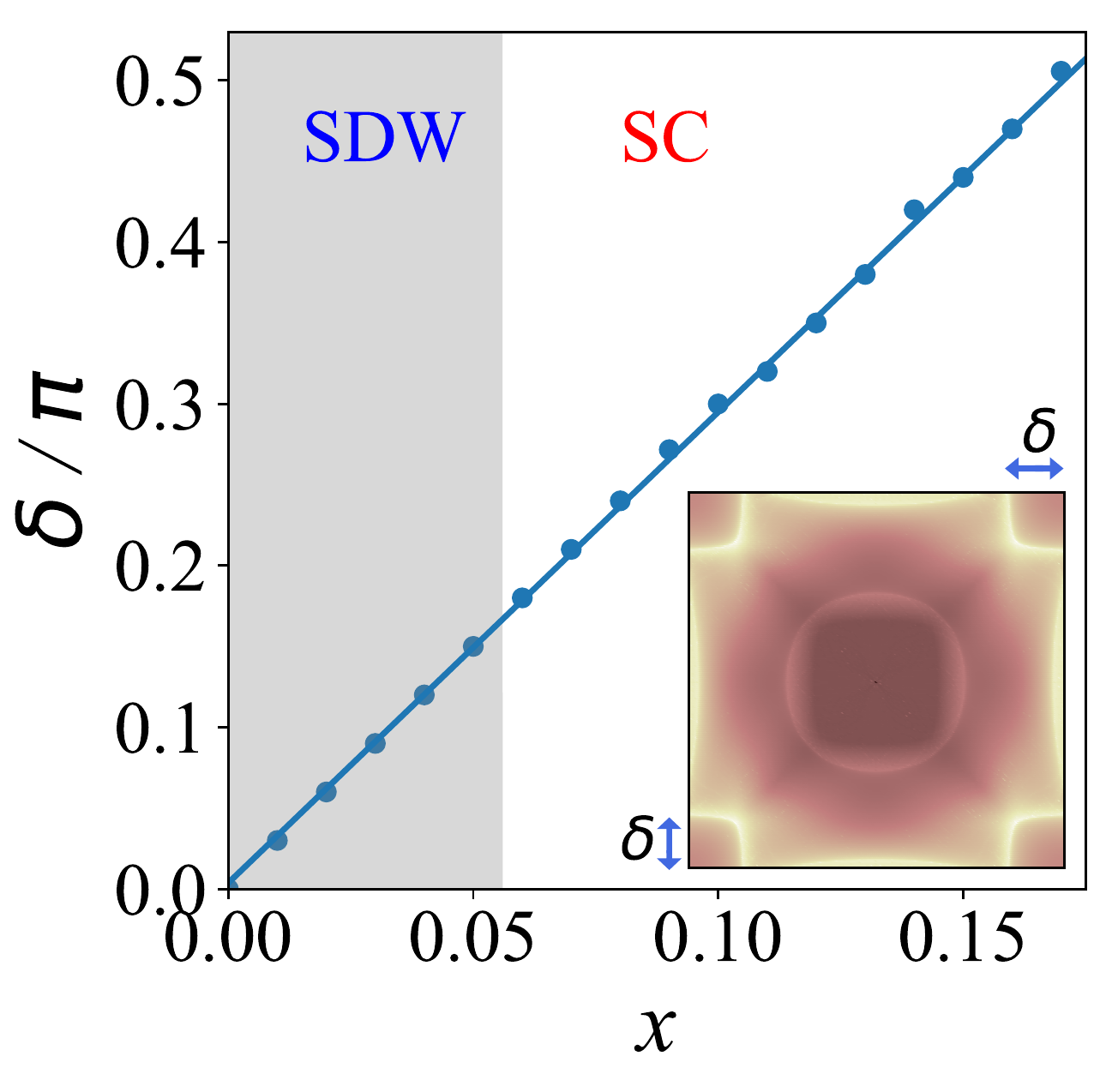}\label{fig:delta}}
	\hspace{0.25in}
	\subfigure[]{\includegraphics[width=1.45in]{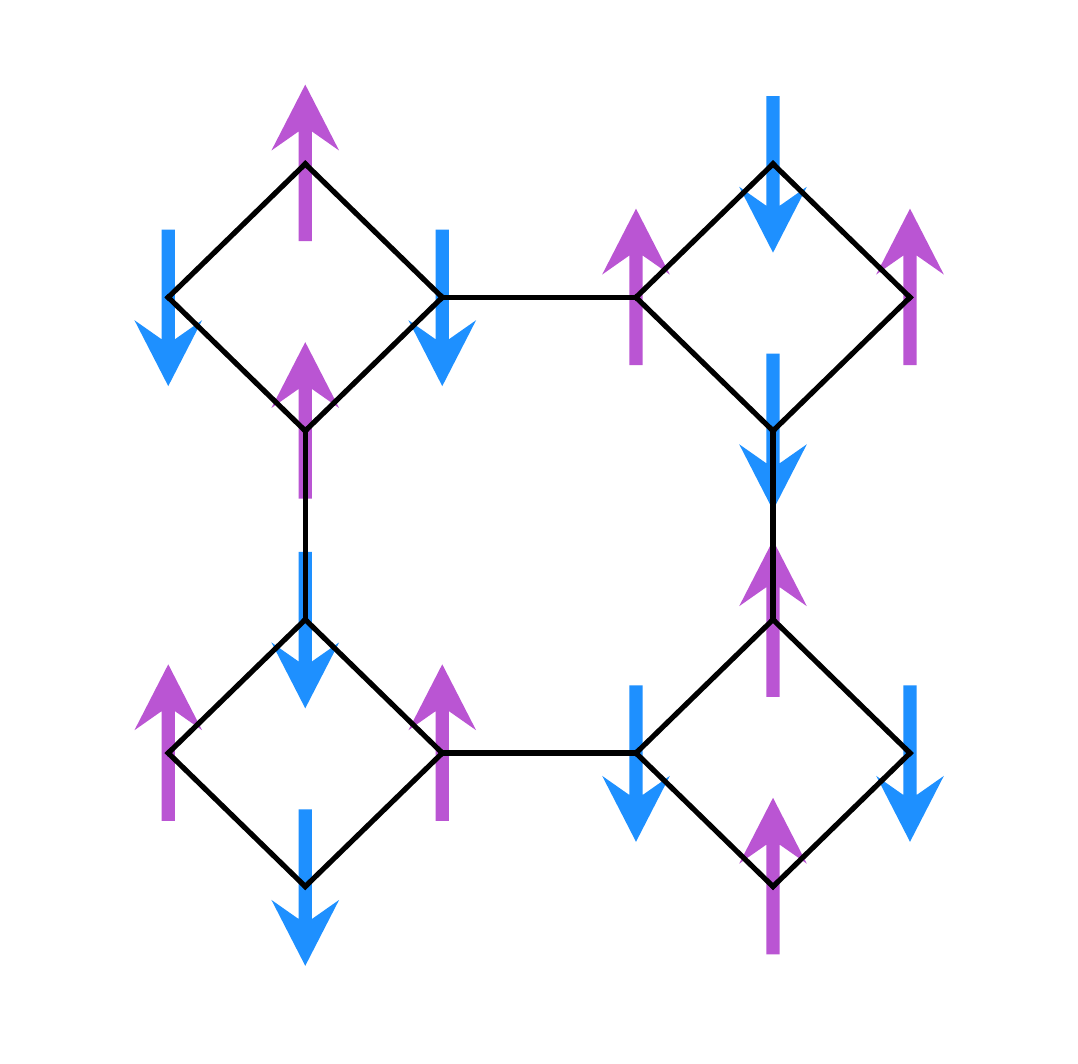}\label{fig:magnetism}}
    \caption{{Panels (a) and (b) show the $\bf{q}$ dependence of the eigensusceptibilities $\chi(\bq)$ in the first Brillioun zone, corresponding to the undoped and $10\%$ electron doped compounds, respectively. The temperature is set as $T = 0.001$.  (c) The incommensurability $\delta$ as a function of doping $x$. (d) The AFM ordered spin pattern in the octagraphene.}
    \label{fig:chi}}
\end{figure}

Let us define the following bare susceptibility for $U=0$:
\begin{align}
    \chi^{(0)l_1l_2}_{l_3l_4} \left(\bq,i\omega_n\right) \equiv
    \frac{1}{N} \int_0^{\beta} d\tau e^{i\omega_n\tau}
    \sum_{\bk_1\bk_2} \big\langle T_{\tau} c^{\dagger}_{l_1}(\bk_1,\tau)          \nonumber \\
    \times c_{l_2}(\bk_1+\bq,\tau)
    c^{\dagger}_{l_3}(\bk_2+\bq,0) c_{l_4}(\bk_2,0) \big\rangle_0.
\end{align}
Here $l_i(i=1,...,4)$ denotes the sublattice indices. The largest eigenvalue $\chi(\bq)$ of the static susceptibility matrix $\chi^{(0)}_{lm}(\bq) \equiv \chi^{(0)l,l}_{m.m}(\bq,i\omega=0)$ for each $\bq$ represents the eigensusceptibility in the strongest channel, while the corresponding eigenvector $\xi(\bq)$ provides information on the fluctuation pattern within the unit cell. The information about the distribution of $\chi(\bq)$ over the Brillouin zone, as well as the fluctuation pattern for the peak momentum, is shown in Fig.~\ref{fig:chi} for different dopings.

Figure \ref{fig:chi0x00} illustrates the distribution of $\chi(\bq)$ over the Brillouin zone for the undoped case, which sharply peaks at $\mathbf{Q}=(\pi,\pi)$, reflecting the perfect FS nesting at that wave vector, as shown in Fig.~\ref{fig:FSn100}. On the other hand, the eigenvector $\xi(\mathbf{Q})=(\frac{1}{2},-\frac{1}{2},\frac{1}{2},-\frac{1}{2})$ reflects the intra-unit-cell fluctuation pattern, which is shown in Fig.~\ref{fig:magnetism} together with the inter-unit-cell pattern for this momentum, which suggests a Neel pattern. With the development of doping, the peak in the distribution of $\chi(\bq)$ splits each into four and deviates from $\mathbf{Q}=(\pi,\pi)$ to $\mathbf{Q}_{x}=(\pi\pm\delta,\pi\pm\delta)$, as shown in Fig.~\ref{fig:chi0x10} for $x=10\%$ electron doping as an example. The relation between $\delta$ and $x$ shown in Fig.~\ref{fig:delta} suggests a linear relation, revealing incommensurate inter-unit-cell fluctuation pattern, just like the Yamada relation in the cuprates\cite{Yamada98}. In the meantime, the eigenvectors $\xi(\mathbf{Q}_{x})$ nearly keep unchanged, and thus the intra-unit-cell fluctuation pattern is still approximately described by Fig.~\ref{fig:magnetism}.

\begin{figure}
	\centering
	\subfigure[]{\includegraphics[height=1.3in]{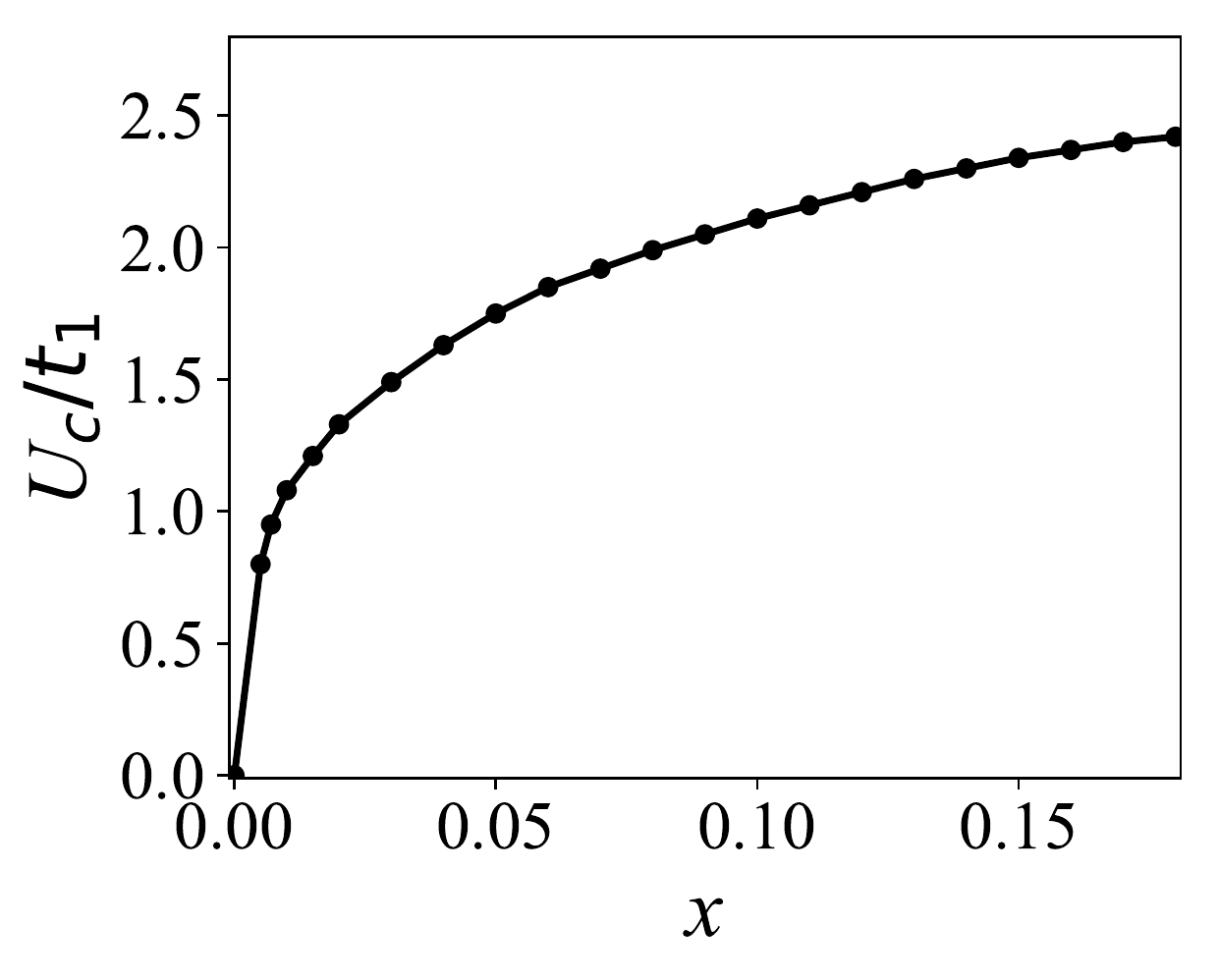}\label{fig:Uc}}
	\subfigure[]{\includegraphics[height=1.3in]{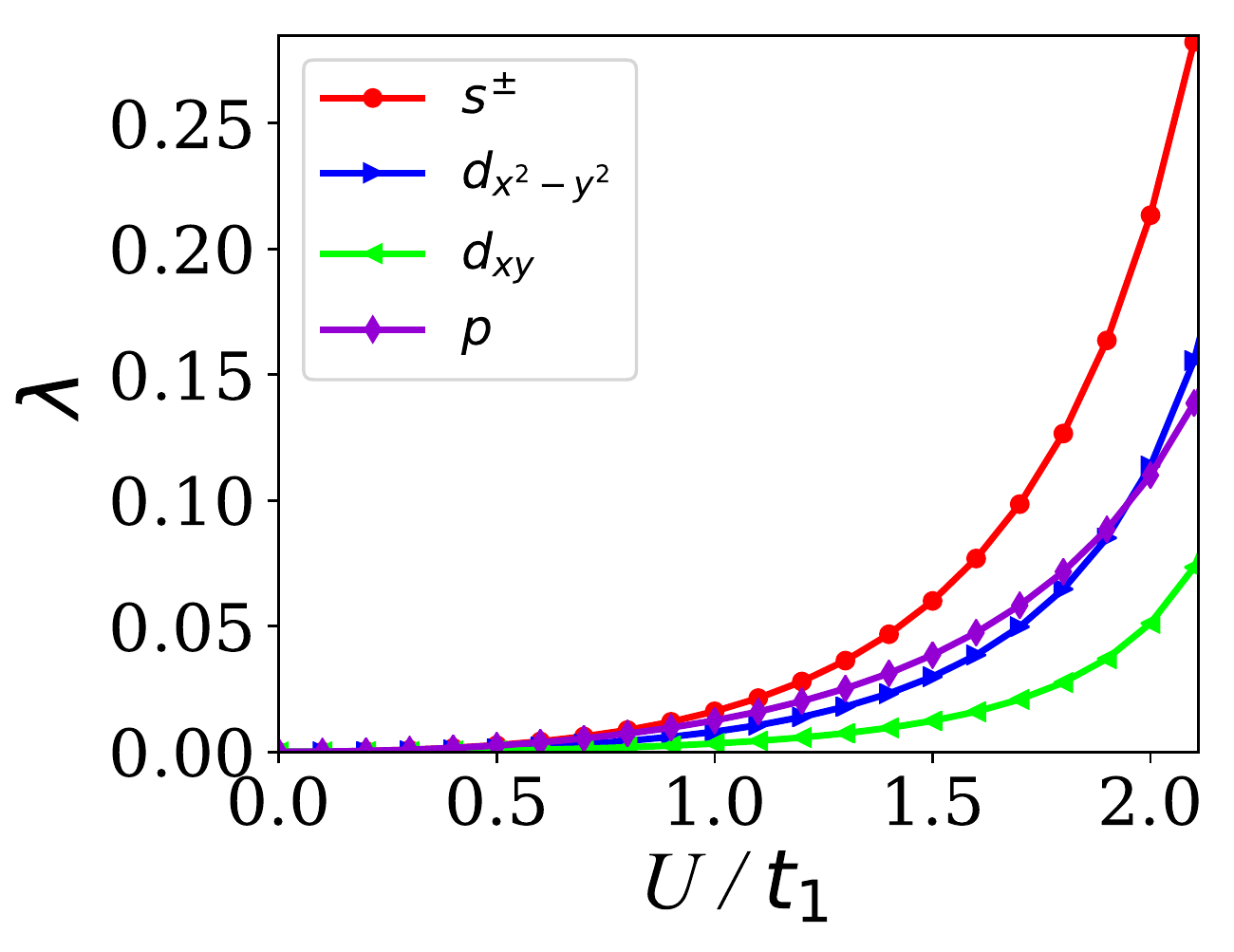}\label{fig:lambda}}
	\\
	\subfigure[]{\includegraphics[height=1.3in]{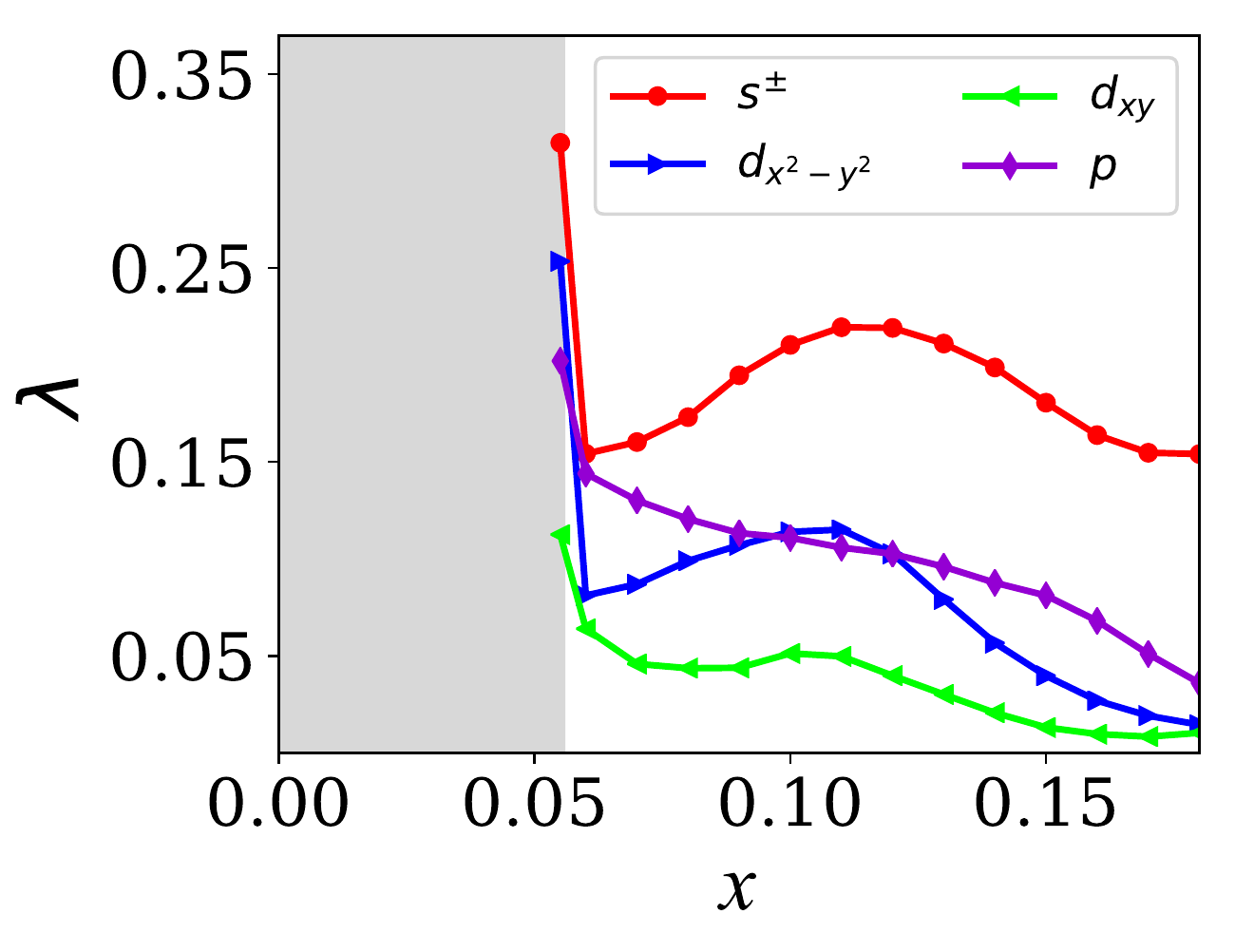}\label{fig:lambda_x}}
	\subfigure[]{\includegraphics[height=1.33in]{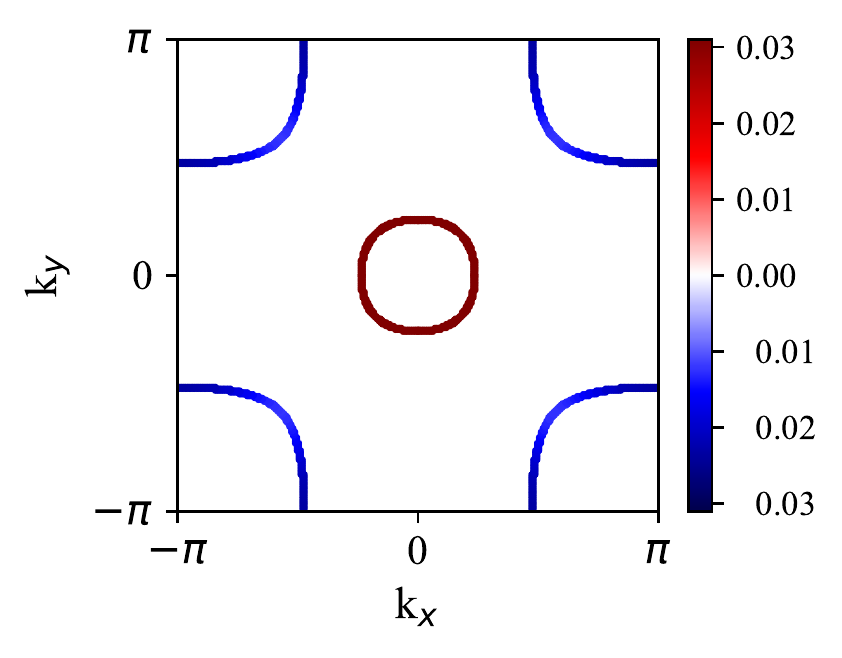}\label{fig:pairing1}}
    \caption{{(a) $U_c/t_1$ as a function of the electron doping density $x$. The largest pairing eigenvalues $\lambda$ in four different pairing symmetry channels as a function of (b) $U/t_1$ and (c) $x$. (d). The {\bf k}-dependent superconducting order parameter $\Delta_{\alpha}({\bk})$ projected onto the FS for the leading $s^\pm$-wave pairing. The doping density for panels (b) and (d) is $x=10\%$. The interaction parameter adopted is $U=1.8t_1$.}}
\end{figure}

For $U>0$, we obtain the following renormalized spin (s) and charge (c) susceptibilities at the RPA level,
\begin{align}
    \chi^{(s/c)}\left(\bq,i\omega_n \right) =
    \left[I \mp \chi^{(0)}\left(\bq,i\omega_n\right)(U)\right]^{-1} \chi^{(0)}\left(\bq,i\omega_n\right)
    \label{eq:RPA}
\end{align}
Here $\chi^{(s/c)}\left(\bq,i\omega_n\right)$, $\chi^{(0)}\left(\bq,i\omega_n\right)$ and $(U)$ are used as $4^2 \times 4^2$ matrices and $I$ is the unit matrix. In our model, $U^{l_1 l_2}_{l_3 l_4} = U \delta_{l_1=l_2=l_3=l_4}$. For $U>0$, the spin fluctuation dominates the charge fluctuation, thus the fluctuation pattern illustrate in Fig.~\ref{fig:magnetism} actually describes the spin fluctuation. Note that the RPA approach only works for $U<U_c$, with the critical interaction strength $U_c$ determined by $\det\left[I - \chi^{(0)}\left(\bq,0\right)U\right]=0$. For $U>U_c$ the spin susceptibility diverges, which suggests that long range SDW order with the pattern shown in Fig.~\ref{fig:magnetism} emerges. The doping-dependence of $U_c$ is shown in Fig.~\ref{fig:Uc}, where one finds $U_c=0$ for $x=0$ due to the perfect FS-nesting, which means that arbitrarily weak repulsive interaction will cause SDW order. For $x>0$, we have $U_c>0$. In such cases, the SDW order maintains for some doping regime where $U_c<U$, but with the wave vector shifting to incommensurate values $\mathbf{Q}_{x}=(\pi\pm\delta,\pi\pm\delta)$.

When the doping concentration $x$ further increases so that $U<U_c$, the long-ranged SDW order is killed. In such parameter regime, the remnant SDW fluctuation will mediate an effective pairing potential $V^{\alpha\beta}(\bk,\bk^\prime)$ \cite{FengLiu13,XXWu15} between the Cooper pairs. Then we can solve the following linearized gap equation to determine the leading pairing symmetry:
\begin{align}
    -\frac{1}{(2\pi)^2}\sum_\beta \oint_{FS} d\bk^\prime_{\parallel} \frac{V^{\alpha\beta}(\bk,\bk^\prime)}{v^\beta_F(\bk^\prime)}
    \Delta_\beta(\bk^\prime) = \lambda\Delta_\alpha(\bk).
    \label{eq:gap}
\end{align}
Here $v^\beta_F(\bk)$ is the Fermi velocity and $\bk^\prime_{\parallel}$ denotes the component along the FS. The pairing eigenvalue $\lambda$ is related to $T_c$ through $T_c\approx W_{D} e^{-1/\lambda}$ with the ``Debye frequency'' $W_D$ for the spin fluctuations to be about an order of magnitude lower than the bandwidth, and the pairing symmetry is determined by the eigenfunction $\Delta_\alpha(\bk)$ corresponding to the largest $\lambda$.

\begin{figure}[htbp]
	\centering
	\subfigure[]{\includegraphics[height=1.25in]{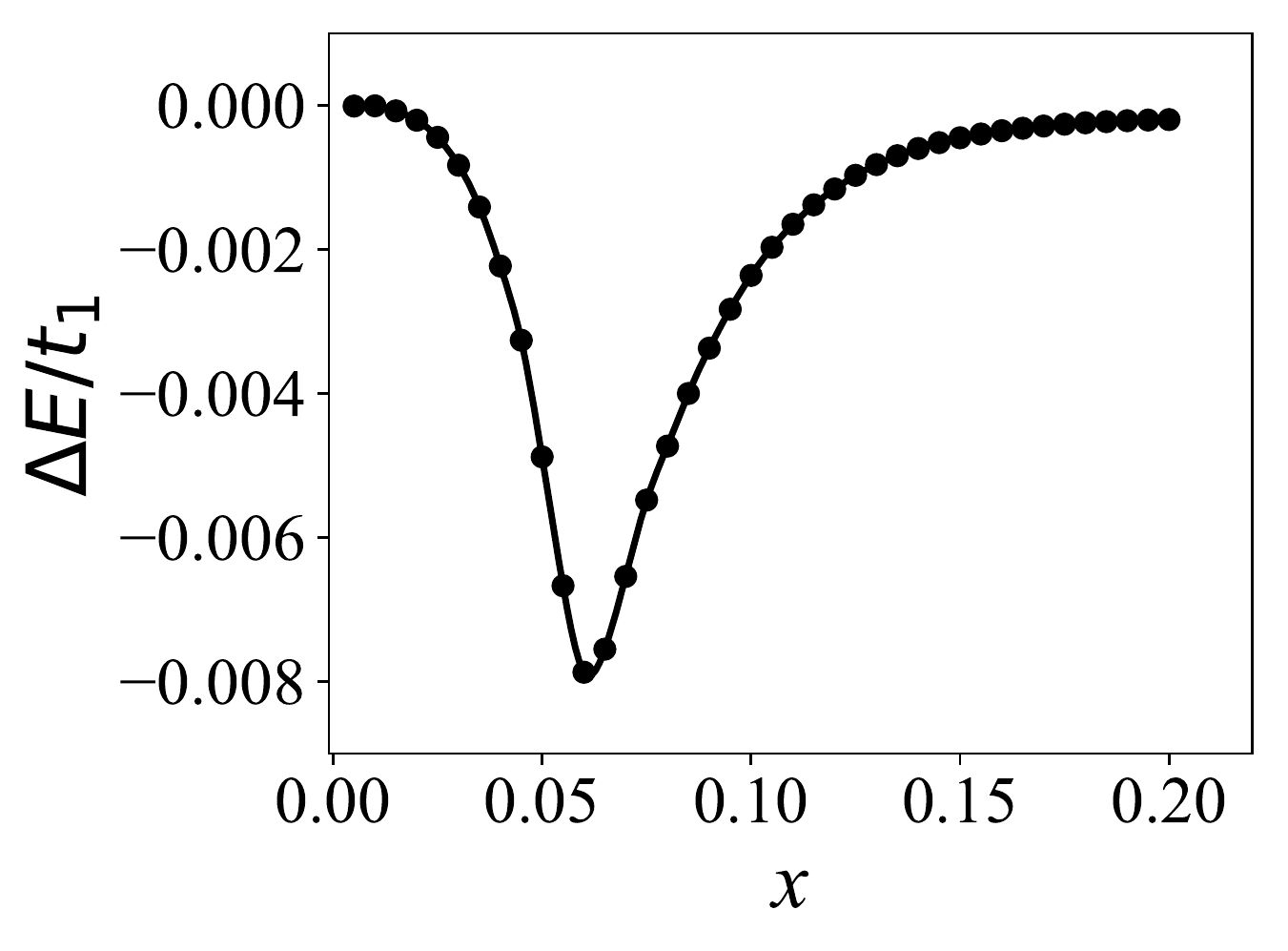}\label{SBMF-a}}
	\hspace{0.05in}
	\subfigure[]{\includegraphics[height=1.25in]{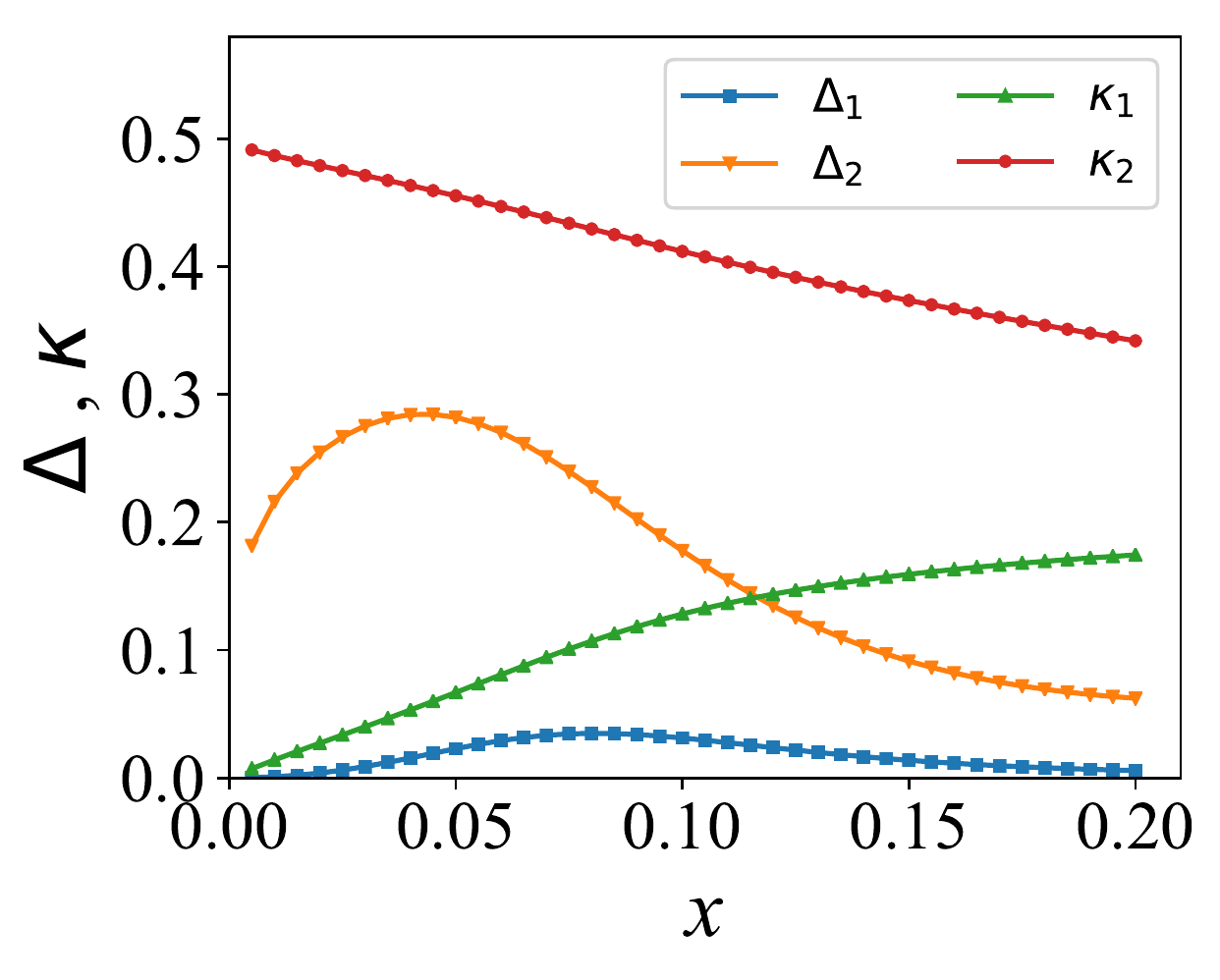}\label{SBMF-b}}
	\\
	\subfigure[]{\includegraphics[height=1.3in]{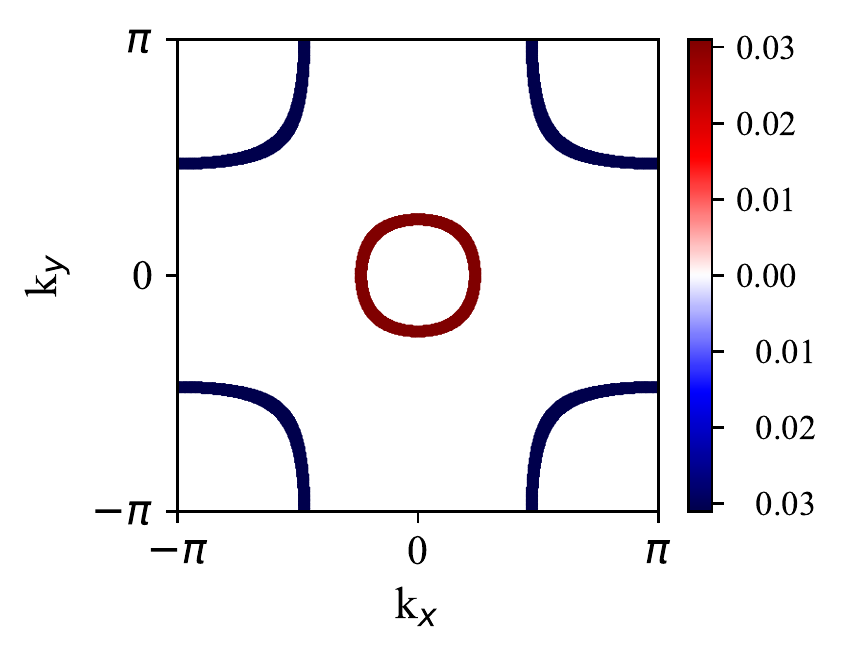}\label{SBMF-c}}
	\subfigure[]{\includegraphics[height=1.25in]{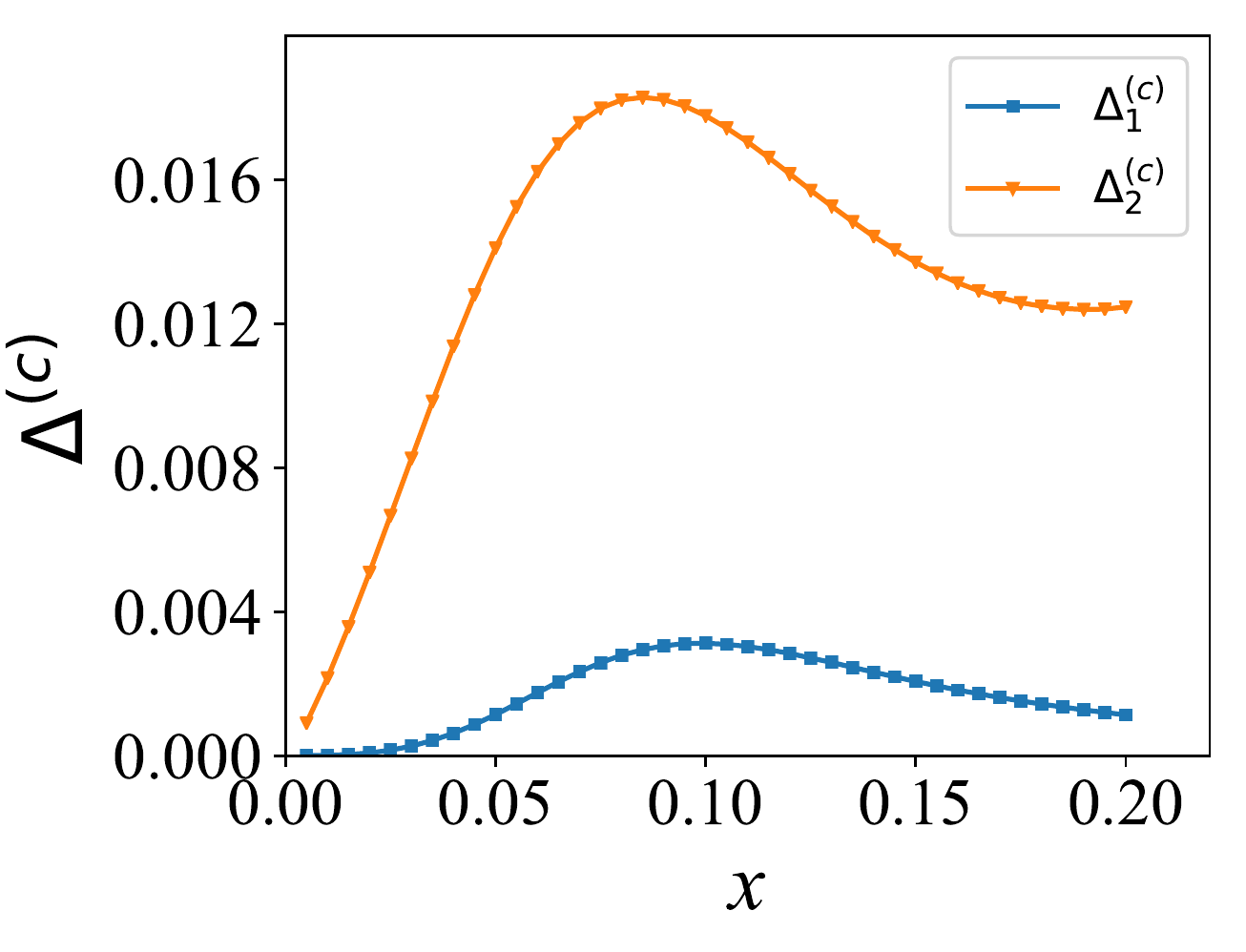}\label{SBMF-d}}
	\caption{(color online). The SBMF results. (a) Doping-dependence of the energy (per unit cell) difference between the $s$-wave pairing and the $d$-wave one, $\Delta E\equiv E_s-E_d$, in units of $t_1$. (b) Doping-dependence of the four SBMF order parameters for the $s$-wave solution. (c) The $s$-wave gap function projected on the FS. (d) Doping dependence of the superconducting order parameter.}
\label{SBMF}
\end{figure}

The $U$ dependence of the largest $\lambda$ for each pairing symmetry is shown in Fig.~\ref{fig:lambda} for a typical doping $x=10\%$.
Obviously, $\lambda$ enhances promptly with the growth of $U$ due to the enhancement of spin fluctuations.
The leading pairing symmetry turns out to be the $s$-wave. In Fig.~\ref{fig:lambda_x}, the doping-dependence of the largest $\lambda$ for each pairing symmetry is shown for a typical $U=1.8t_1$. After a prompt drop near the critical doping (about $\pm5\%$), the $\lambda$ for the four pairing symmetries vary smoothly for a wide doping range up to $20\%$, where the $s$-wave SC dominates all the other pairings. Figure \ref{fig:lambda} and \ref{fig:lambda_x} illustrates the robustness of the $s$-wave SC against parameters variation. The $C_{4v}$-symmetric distribution of the pairing gap function $\Delta({\bk})$ of the obtained $s$-wave SC is shown on the FS in Fig.~\ref{fig:pairing1}. Remarkably, this gap function keeps the same sign within each pocket and changes sign between the two pockets. Therefore, we have established here a one-orbital realization of the standard $s^\pm$ SC, which used to be realized in the multi-orbital Fe-based superconductor family.

Note that the interaction parameter $U=1.8t_1\approx4.5$ eV adopted here is considerably weaker than realistic value of $U\approx10$ eV \cite{Neto}, and due to the weak-coupling perturbative character of RPA, it is unreasonable to adopt a stronger $U$. In the next section, we adopt the SBMF approach to treat with the strong-coupling limit.

\subsection{The slave-boson mean-field results}
We start from the following effective $t$-$J$ model to study the strong-coupling limit of the Hubbard-model (\ref{model}),
\begin{align}
H=H_{\text{TB}}+J_1\sum _{\left \langle i,j \right \rangle  }\bm{\widehat{S}_i}\cdot \bm{\widehat{S} } _{j}   +J_2\sum _{\left [ i,j \right ]  }\bm{\widehat{S} }_{i}\cdot \bm{\widehat{S} } _{j},\label{tJ}
\end{align}
Here the intrasquare $NN$ ($J_{1}$) and intersquare $NN$ ($J_{2}$) effective superexchange coupling constants are generated in the strong-coupling limit, which roughly satisfy $J_2/J_1\approx(t_2/t_1)^2\approx1.4$. In the following, we adopt $J_1=0.5t_1$ and $J_2=0.7t_1$. This Hamiltonian should be understood as acting on the subspace of empty (double-occupancy) and  single occupied sites for the hole-doped (electron-doped)  system.

In the SBMF approach\cite{Kotliar}, we decompose the electron operator $c_{i\sigma}$ into $c_{i\sigma}\to f_{i\sigma}b^{\dagger}_i$, with the bosonic holon (doublon) operator $b^{\dagger}_i$ and the fermionic spinon operator $f_{i\sigma}$ subject to the no-double-occupancy constraint $b^{\dagger}_ib_i+\sum_{\sigma}f^{\dagger}_{i\sigma}f_{i\sigma}=1$. This constraint is treated in the mean-field level in SBMF, and at zero temperature the condensation of bosonic $b^{\dagger}_i$ leads to $b^{\dagger}_i\to \sqrt{x}$ and we are left with only the fermionic $f_{i\sigma}$ degree of freedom. The quartic term of $f_{i\sigma}$ in $H$ is further mean-field decomposed with the following two order parameter channels:
\begin{align}\label{sce}
\kappa _{(i,j)} =&\left \langle  f^{\dagger}_{j\uparrow }f_{i\uparrow} \right \rangle=\left \langle  f^{\dagger}_{j\downarrow }f_{i\downarrow} \right \rangle \nonumber\\
\Delta _{(i,j)}=&\left \langle  f_{j\downarrow }f_{i\uparrow}-f_{j\uparrow }f_{i\downarrow}\right \rangle.
\end{align}
Here we actually have two mean-field $\kappa _{(i,j)}$ ($\Delta _{(i,j)}$) parameters, i.e., $\kappa _{1}$ ($\Delta _{1}$) for intrasquare $NN$ and $\kappa _{2}$ ($\Delta _{2}$) for intersquare $NN$ $(i,j)$, respectively, which are obtained by solving the mean-field equation self-consistently.

Our SBMF results are shown in Fig.~\ref{SBMF}. Here we have tried two different pairing symmetries, i.e., the $s$ wave and $d$ wave, with their total energy difference $\Delta E\equiv E_s-E_d$ shown in Fig.~\ref{SBMF-a}, where the $s$-wave SC gains more energy and becomes the ground state. The doping dependence of the four order parameters $\kappa _{1,2}$ and $\Delta _{1,2}$ for the $s$-wave pairing is shown in Fig.~\ref{SBMF-b}, where the intersquare order parameters obviously dominate the intrasquare ones. Figure \ref{SBMF-c} shows the projection of the gap function onto the FS, where one clearly verifies the standard $s^\pm$-pairing state, which is well consistent with the gap function obtained by RPA shown in Fig.~\ref{fig:pairing1}.

The doping-dependence of the superconducting order parameter $\Delta^{(c)} _{(i,j)}=\left \langle  c_{j\downarrow }c_{i\uparrow}-c_{j\uparrow }c_{i\downarrow}\right \rangle=x \Delta _{(i,j)}$ is shown in Fig.~\ref{SBMF-d}, which illustrates a dome-shape similar to the cuprates. If we use the BCS relation $2J\Delta^{(c)}/T_c\approx3.53$ to roughly estimate $T_c$, we get the highest $T_c\approx180$ K near $x=10\%$ for our choice of $J_1$ and $J_2$. However, as the effective superexchange parameters $J_1$ and $J_2$ for real material with intermediate $U$ is hard to estimate, the $T_c$ obtained here might not be accurate. In the following, we adopt the VMC approach to study the problem.

\subsection{The variational Monte Carlo results}
The above weak-coupling RPA and strong-coupling SBMF approaches consistently yield the $s^\pm$-wave pairing. However, to obtain a more reasonable estimation of $T_c$, we should adopt a realistic interaction parameter $U$. The realistic $U\approx10$ eV is comparable with the total bandwidth, thus it belongs to intermediate coupling strength. We adopt the VMC approach here, which is suitable for the intermediate coupling strength.

We adopt the following partially Gutzwiller-projected BCS wave function \cite{YangVMC} in our VMC study,
\begin{align}\label{wave}
\left  |G \right \rangle=g^{\sum_{i} n_{i\uparrow}n_{i\downarrow}} (\sum_{\bf{k}\alpha}\frac{v^{\alpha }_{\bm{k}}}{u^{\alpha }_{\bm{k}}}
 c^{\dagger}_{\bf{k}\alpha\uparrow}c^{\dagger}_{\bf{-k}\alpha \downarrow})^{\frac{N_e}{2}}\left  |0 \right \rangle.
\end{align}
Here $g\in(0,1)$ is the penalty factor of the double occupancy, $N_e$ is the total number of electrons, and 
\begin{align*}
\frac{v^{\alpha }_{\bm{k}}}{u^{\alpha }_{\bm{k}}}=\frac{\Delta^{\alpha}_{\bm{k}}}{\varepsilon_{\alpha }(\bm{k})+\sqrt{\varepsilon^2_{\alpha }(\bm{k})+\left |\Delta ^{\alpha }_{\bm{k}}\right |^{2} }},
\end{align*}
where $\Delta ^{\alpha }_{\bm{k}}=\Delta ^{\alpha }f(\bm{k})$ is the superconducting gap function. Here we only consider intra-band pairing on the $\alpha=2,3$ bands crossing the FS, with $\Delta^{2}=\Delta^{3}\equiv\Delta$. The following four different form factors $f(\bm{k})$ are considered in our calculations,
\begin{align}\label{factor}
f(\bm{k})=\left\{\begin{matrix}
\cos k_{x}+\cos k_{y}\quad &(s^\pm)\\
\cos k_{x}\cos k_{y}\quad &(s^{++})\\
\cos k_{x}-\cos k_{y}\quad &(d_{x^{2} -y^{2}})\\
\sin k_{x}\sin k_{y}\quad &(d_{xy})
\end{matrix}\right.
\end{align}
There are three variational parameters, i.e., $g$, $\mu_c$, and $\Delta$ for each pairing channel in our trial wave function.

\begin{figure}
    \subfigure[]{\includegraphics[width=1.68in]{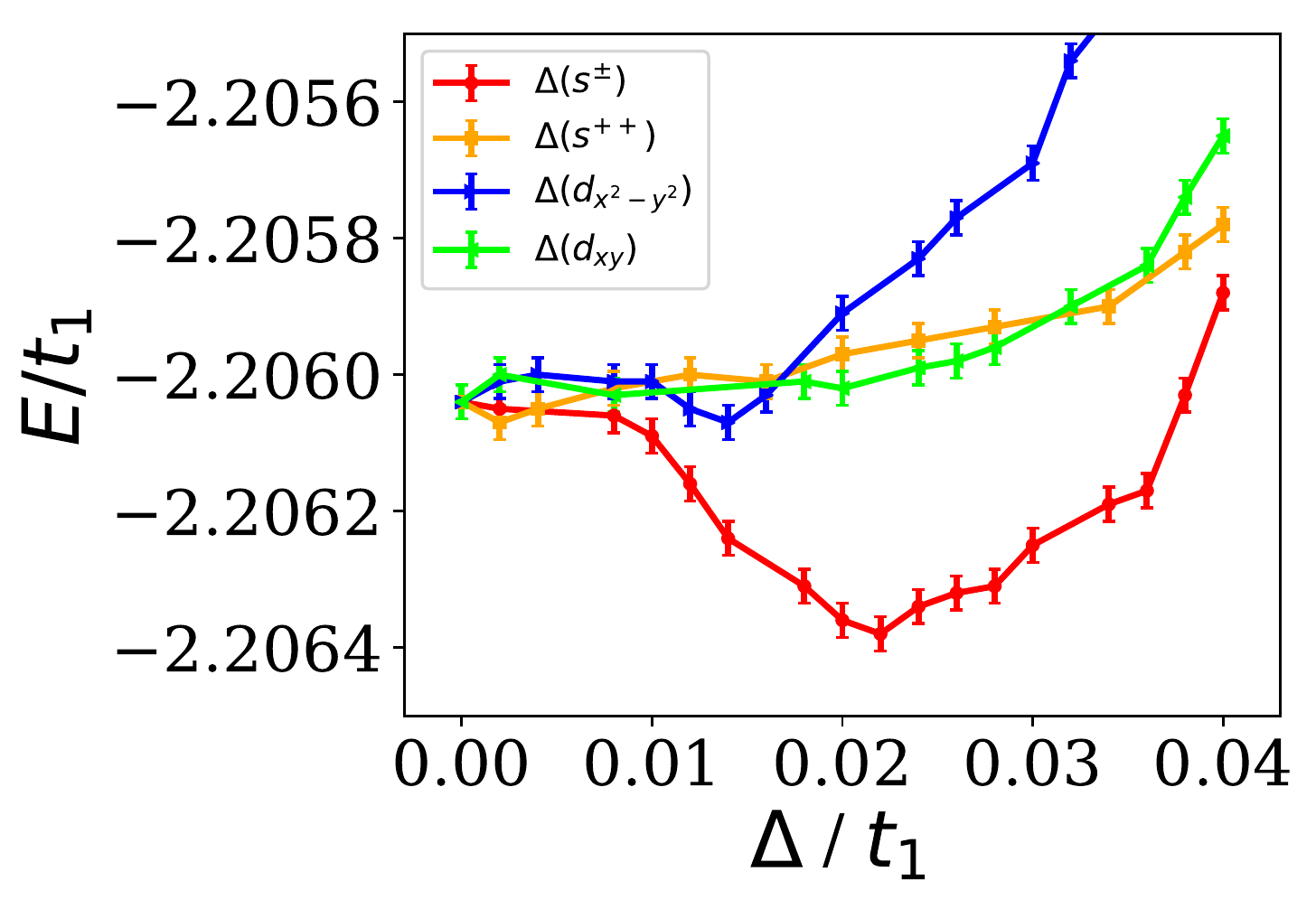}\label{ttu}}
    \subfigure[]{\includegraphics[width=1.68in]{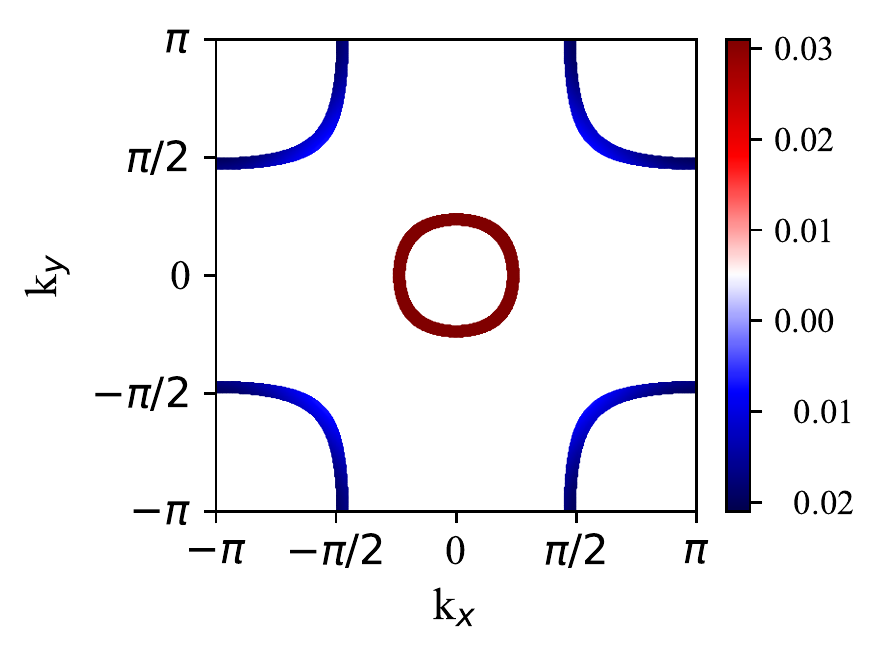}\label{ttmu}}
    \caption{(a) The VMC results for the energy per unit cell as function of $\Delta$ for the four different gap form factors $s^\pm$, $s^{++}$, $d_{x^{2} -y^{2}}$ and $d_{xy}$ with $g$ and $\mu_c$ optimized for each $\Delta$.  (b) The $\bm{k}-$dependent superconducting order parameter $\Delta(\bm{k})$ projected on the FS for the 10\% electron-doped compound. The interaction parameter adopted is $U=4t_1=10$ eV.}
\end{figure}

We employ the VMC approach to calculate the expectation value $E$ of the Hubbard Hamiltonian (\ref{model}) \cite{YangVMC} and optimize the variational parameters. The $\Delta$ dependence of the energy per unit cell for each form factor is shown in Fig.~\ref{ttu} for $U=4t_1=10$eV for a typical doping $x=10\%$, with $g$ and $\mu_c$ optimized for each $\Delta$. Note that the optimized $g=0.5475$ is almost equal to the optimized value without SC, and that $\mu_c$ is almost equal to the value obtained in the mean-field calculation. From Fig.~\ref{ttu}, one finds that the $s^\pm$- wave pairing causes the most energy gain among the four gap form factors, with the optimized gap amplitude at $\Delta=0.022t_1\approx$50meV, comparable with the cuprates, implying similar $T_c$ between them. The gap function of the $s^\pm$-wave SC obtained is shown on the FS in Fig.~\ref{ttmu}, which is well consistent with that obtained in the RPA calculation.

Note that we have not included antiferromagnetic order in our trial wave function as we mainly focus on SC here. Generally, such antiferromagnetic order will be favored at low dopings and decay with further doping. In the framework of VMC, the antiferromagnetic order possibly coexists with SC at low dopings. We leave this topic for future studies.

\section{Discussion and Conclusion}
The synthesis of octagraphene is on the way. Recently, graphene-like nanoribbons periodically embedded with four- and eight-membered rings have been synthesized \cite{Zhong}. A scanning tunneling microscopy and atomic force microscopy study revealed that four- and eight-membered rings are formed between adjacent perylene backbones with a planar configuration. This 2D material can be taken as an intermediate between the graphene and the octagraphene studied here. Most probably, the octagraphene might be synthesized in the near future, which will provide a material basis for the study here.

In conclusion, we have studied possible pairing states in the single-orbital Hubbard model on the square-octagon lattice with only nearest-neighbor hopping terms. Due to the perfect FS nesting in the undoped system, slight doping would induce HTCS, driven by strong incommensurate SDW fluctuations. Our combined RPA-, SBMF-, and VMC-based calculations suitable for the weak, strong, and intermediate couplings strengths, respectively, consistently yield standard $s^\pm$-wave SC in this simple one-orbital system. The smoking-gun evidence of this intriguing pairing state would be the pronounced subgap spin resonance mode emerging upon the superconducting transition, which can be detected by inelastic neutron scattering. We propose octagraphene as a possible material realization of the model, and our VMC calculations adopting realistic interaction parameter for this material yield a pairing gap amplitude of about 50 meV, comparable with that of the cuprates, which implies comparable $T_c$ between the two systems. Our study will also apply to other materials with similar lattice structure. Our results, if confirmed, would start a new stage in the discovery of high-$T_c$ SC.

\begin{acknowledgments}
F.Y. acknowledges the support from NSFC under the
Grants No. 11674025, No. 11334012, and No. 11274041.
Y.-T.K. and D.-X.Y. are supported by NKRDPC Grants No.
2017YFA0206203, No. 2018YFA0306001, NSFC-11574404,
and NSFG-2015A030313176, Special Program for Applied
Research on Super Computation of the NSFC-Guangdong
Joint Fund, National Supercomputer Center In Guangzhou,
and Leading Talent Program of Guangdong Special Projects.
\end{acknowledgments}

\null\vskip-8mm

\bibliographystyle{apsrev4-1}
\bibliography{RPA}

\end{document}